\documentclass[bm,aps,showpacs,amsmath,preprint,amssymb,12pt]{revtex4}
\usepackage{graphicx}
\usepackage{bm}

\begin{document}
{~}
\title{Charged Rotating Kaluza-Klein Black Holes \\Generated by $G_{2(2)}$ Transformation}

\vspace{2cm}
\author{Shinya Tomizawa$^{1}$, Yukinori Yasui$^{2}$ and Yoshiyuki Morisawa$^{3}$}

\vspace{2cm}
\affiliation{
${}^{1}$Cosmophysics Group, Institute of Particle and Nuclear Studies, 
KEK, Tsukuba, Ibaraki, 305-0801, Japan \\
${}^{2}$Department of Mathematics and Physics,Graduate School of Science,Osaka City University, 3-3-138 Sugimoto, Sumiyoshi, Osaka 558-8585, Japan\\
${}^{3}$Faculty of Liberal Arts and Sciences, Osaka University of Economics and Law, Yao City, Osaka 581-8511, Japan}

\begin{abstract}
Applying the {\it $G_{2(2)}$ generating technique for minimal $D=5$ supergravity} to the Rasheed black hole solution, we present a new rotating charged Kaluza-Klein black hole solution to the five-dimensional Einstein-Maxwell-Chern-Simons equations.  At infinity, our solution behaves as a four-dimensional flat spacetime with a compact extra dimension and hence describes a Kaluza-Klein black hole. In particlar, the extreme solution is non-supersymmetric, which is contrast to a static case. Our solution has the limits to the asymptotically flat charged rotating black hole solution and a new charged rotating black string solution.
\end{abstract}

\preprint{KEK-TH 1276\ OCU-PHYS 304}
\date{\today}
\maketitle

\section{Introduction}

In recent years classical general relativity in more than four spacetime dimensions has been the subject of increasing attention. In particular, higher dimensional black objects have become one of 
the major subjects in fundamental physics~\cite{MP,ER,MI,PS,F,IM_BD,EF,BMPV,SUSYRing,SUSYRing2,C_ring,C_ring2,CLP,Cai,Helfgott,Galloway,Morisawa,MTY,HIW}.  
Although recent 
ideas of braneworld and TeV gravity~\cite{ADD,RS1,RS2} 
have opened up the possibility of large extra-dimensions, still important 
and widely believed especially in the context of string theories 
is the spacetime picture that our macroscopically large four-dimensional 
spacetime is realized from a higher dimensional 
spacetime by some mechanism of compactifying  
extra-dimensions within the size of the fundamental scale of gravity. 
In this context, it is of great interest and important to study Kaluza-Klein black holes, which are essentially higher dimensional near the event horizon but effetively looks like four-dimensional at infinty.    
Also, in the context of string theory, the five-dimensional
Einstein-Maxwell theory with a Chern-Simon term gathers
much attention since it is the bosonic sector of the minimal
supergravity. Supersymmetric  black hole solutions
to the five-dimensional Einstein-Maxwell equations with a Chern-Simons terms have been found by various authors.

\medskip 

In this context, one of recent important works is the discovery of Ishihara-Matsuno solution~\cite{IM}, which is a static charged Kaluza-Klein black hole solution in the five-dimensional Einstein-Maxwell theory. They found it by using the squashing transformation, i.e., they regarded $\rm S^3$ of a five-dimensional 
Reissner-Nordst\"om black hole as a twisted $\rm S^1$ bundle over $\rm S^2$ and  changed the ratio of 
the radius of $\rm S^2$ to that of $\rm S^1$ fiber such that the ratio diverges at infinity. A static charged multi Kaluza-Klein 
black hole solution~\cite{IKMT} was constructed immediately from 
the Ishihara-Matsuno solution. Subsequently the squashing transformation were used to generate many different Kaluza-Klein black hole solutios. In fact, it was demonstrated by Wang~\cite{Wang} that the five-dimensional Kaluza-Klein black hole of Dobiasch and Maison~\cite{DM,GW} can 
be reproduced by squashing a five-dimensional Myers-Perry black hole with two equal angular momenta.    
In \cite{NIMT}, the squashing transformation was also applied 
to the Cvetic {\it et al.}'s charged rotating black hole 
solution~\cite{CLP} with two equal angular momenta to the five-dimensional Einstein-Maxwell-Chern-Simons equations. Moerover, the application of the squashing transformation to non-asymptotically 
flat Kerr-G\"odel black hole solutions~\cite{Gimon-Hashimoto} was 
considered in \cite{TIMN,MINT,TI}.  
In particular, the solution \cite{NIMT} is the  
a generalization of the Ishihara-Matsuno solution to the rotating 
black holes in Einstein-Maxwell-Chern-Simons theories, which we will comare our new solution in this paper . The solution in general  describes a non-supersymmetric black hole boosted along the direction of the extra dimension and has the limit to the rotating supersymmetric black hole solution.
A similar type of Kaluza-Klein black holes was considered in the context of 
supersymmetric theories by Gaiotto {\it et al.}~\cite{Gaiotto} 
and Elvang {\it et al.}~\cite{EEMH}.  
Furthermore, Kaluza-Klein solutions of this type are generalized to various different solutions by many authors~\cite{BKW,HO,KRT,IKMT2,IKT,MIKT,TIKM,T,IIKMMT,YIKMT}.

\medskip 
As mentioned in~\cite{TI}, however, the squashing transformation can be applied only to a special class of solutions, i.e., cohomogeneity-one black hole solutions such as the five-dimensional Myers-Perry black hole solution~\cite{MP} with two equal angular momenta and Cvetic-L\"u-Pop black hole solution~\cite{CLP} with two equal angular momenta. New solutions generated by the squashing transformation also belongs to a class of cohomogeneity-one. Therefore, 
to obtain more general cohomogeneity-two spacetimes such as  Kaluza-Klein black hole with two rotations, or to applying it to cohomogeneity-two solutions such as black rings~\cite{ER,PS}, black holes with two rotations~\cite{MP}, black lens~\cite{LMP,LMP2}, one has to consider a generalization of the squashing 
technique, or quitely different methods.

\medskip 
Recently, a new solution-generation technique for $D=5$ minimal supergravity have been developed by Bouchareb {\it et al.}~\cite{BCCGSW}. Under the assumptions of the exsistence of two commuting Killing vector fields, the five-dimensional Einstein-Maxwell-Chern-Simons equations reduces to equations for eight scalar fields on a three dimensional space. As first discussed by Mizoguchi {\it et al.}~\cite{MO,MO2}, the system of the scalar fields is described by a non-linear sigma model whose action is invariant under $G_{2(2)}$ transformations. In particular, the target space of the three-dimensional sigma model is the coset $G_{2(2)}/[SL(2,{\bm R})\times SL(2,{\bm R})]$ in the Euclidean case. Some isometries of the target space construct solution-generation symmeties. They have constructed a simple representation of a coset in terms of $7\times 7$ matrices. To obtain a new solution from a seed solution, one need transform the coset matrices for the seed by one-parameteric subgroups of $G_{2(2)}$. In fact, using the formalism, they succeeded in derivation of charged black hole solution~\cite{CLP} from the five-dimensional Myers-Perry black hole solution~\cite{MP}.

\medskip 
Our purposes of this paper are to construct a new charged rotating Kalzua-Klein black hole solution which belongs to a class of cohomogeneity-two in the five-dimensional Einstein-Maxwell-Chern-Simons theory by using the $G_{2(2)}$ transformation from the Rasheed solution~\cite{Rasheed} and to examine several features of the solution. Our solution can be regarded as the generalization of the Ishihara-Matsuno solution to the rotating case. The black hole is rotating only in the four dimensional direction, which is contrast to the charged rotating Kalzua-Klein black hole solution in~\cite{NIMT} generated by the squashing transformation. Namely, the black hole is rotating only along fifth dimensional derection. As shown later, these two black hole solutions have some different properties. In particular, it is interesting to study the difference between two solutions in the extreme cases in the context of supersymmetric theoty. In that case, the solution in~\cite{NIMT} can become supersymmetric but our solution are non-supersymmetric. Our new solution has many limits to several known solutions and a new electrically charged rotating black string solution.

\medskip 
The rest of this paper is organized as follows. 
In the next section, we will shortly review the solution-generation techniques based on the results of the paper~\cite{BCCGSW}. 
Then, in section \ref{sec:Rasheed}, we review the Rasheed solution, which is considered as a seed solution in our work. 
In section \ref{sec:new}, following the results in Ref.\cite{BCCGSW}, we will generate a new solutions from the Rasheed solution under several special 
choices of the parameters and in section \ref{sec:property}, we will study their basic properties. 
Section \ref{sec:summary} is devoted to summary and discussions.

\section{$G_{2(2)}$ generation technique }
Here, following Ref.~\cite{BCCGSW}, we give the results of the $G_{2(2)}$ transformation which generates charged solutions to the five-dimensional Einstein-Maxwell-Chern-Simons equations from solutions to the five-dimensional vacuum Einstein equations.

\medskip
Consider the five-dimensional Einstein-Maxwell theory with a Chern-Simons  
term, whose action~\footnote{The actions written here and in \cite{IM,TIMN,MINT,TI}  are diffrent in the coefficients of $F_{\mu\nu}$ and $A_\mu F_{\nu\rho}F_{\sigma\lambda}$ by the multiplier of $1/4$. To avoid confusion, in this paper we follow Ref.\cite{BCCGSW}} is given by  
\begin{eqnarray}
 S = \frac{1}{16 \pi G_5} 
    \int d^5 x \left[ 
                    \sqrt{-g} \left( R - \frac{1}{4}F_{\mu \nu } F^{\mu \nu } \right) 
                    + \frac{1}{12 \sqrt 3} 
                      \epsilon ^{\mu \nu \rho \sigma \lambda } 
                      A_\mu F_{\nu \rho } F_{\sigma \lambda } 
               \right] \,, 
\label{action}
\end{eqnarray}
where $R$ is the five dimensional scalar curvature, 
$F = dA$ is the two-form of the five-dimensional gauge field 
associated with the gauge potential one-form $A$, and $G_5$ is 
the five-dimensional Newton constant. Varying the action (\ref{action}), 
we derive the Einstein equation 
\begin{eqnarray}
 R_{\mu \nu } -\frac{1}{2} R g_{\mu \nu } 
 = \frac{1}{2} \left( F_{\mu \lambda } F_\nu^{ ~ \lambda } 
  - \frac{1}{4} g_{\mu \nu } F_{\rho \sigma } F^{\rho \sigma } \right) \,, 
 \label{Eineq}
\end{eqnarray}
and the Maxwell equation 
\begin{eqnarray}
 F^{\mu \nu}_{~~~; \nu} + \frac{1}{4 \sqrt 3} \left( \sqrt{-g} \right)^{-1} 
   \epsilon ^{\mu \nu \rho \sigma \lambda } F_{\nu \rho } F_{\sigma \lambda } 
   = 0 \,. 
 \label{Maxeq}  
\end{eqnarray}

We assume that the space-time admits two commuting Killing vector fields $\partial_{z_a}\ (a=0,1)$. The metric and the gauge potential one-form can be written as follows, respectively 
\begin{eqnarray}
ds^2=\lambda_{ab}(dz^a+a^a{}_idx^i)(dz^b+a^b{}_jdx^j)+\tau^{-1}h_{ij}dx^idx^j,
\end{eqnarray}
\begin{eqnarray}
{\bm A}=\sqrt{3}\psi_adz^a+A_idx^i,
\end{eqnarray}
where $\tau=-{\rm det}(\lambda_{ab})$ and $i=2,3,4$. Dualization of $F_{ij}=\partial_iA_j-\partial_jA_i$ gives the scalar field $\mu$ by the equation
\begin{eqnarray}
\frac{1}{\sqrt{3}}F^{ij}=a^{aj}\partial^i\psi_a-a^{ai}\partial^j\psi_a+\frac{1}{\tau \sqrt{h}}\epsilon^{ijk}(\partial_k\mu+\epsilon^{ab}\psi_a\partial_k\psi_b),\label{eq:dual1}
\end{eqnarray}
and dualization of $G^b_{ij}=\partial_ia^b_j-\partial_ja^b_i$ gives the two vecors $\omega^a$ by the equation
\begin{eqnarray}
\tau\lambda_{ab}G^{bij}=\frac{1}{\sqrt{h}}\epsilon^{ijk}[\partial_k\omega_a-\psi_a(3\partial_k\mu+\epsilon^{bc}\psi_b \partial_k \psi_c)],\label{eq:dual2}
\end{eqnarray}
where $h={\rm det}(h_{ij})$.

\medskip 
Introduce the $G_{2(2)}/(SL(2,{\bm R})\times SL(2,{\bm R}))$ coset matrix $M$ constructed by
\begin{eqnarray}
M= \left(
  \begin{array}{ccc}
  A&B&\sqrt{2}U\\
  B^T&C&\sqrt{2}V\\
  \sqrt{2}U^T&\sqrt{2}V^T&S\\
  \end{array}
 \right),
\end{eqnarray}
with
\begin{eqnarray}
&&A=\left(
  \begin{array}{ccc}
  [(1-y)\lambda+(2+x)\psi \psi^T-\tau^{-1}\tilde\omega\tilde\omega^T+\mu(\psi \psi^T\lambda^{-1}J-J\lambda^{-1}\psi\psi^T)]&\tau^{-1}\tilde\omega\\
 \tau^{-1}\tilde\omega^T& -\tau^{-1}
  \end{array}
 \right),\nonumber\\
&&B=\left(
  \begin{array}{ccc}
  (\psi\psi^T-\mu J)\lambda^{-1}-\tau^{-1}\tilde\omega \psi^T J&[(-(1+y)\lambda J-(2+x)\mu+\psi^T\lambda^{-1}\tilde\omega)\psi+(z-\mu J\lambda^{-1}\tilde)\omega] \\
  \tau^{-1}\psi^T J&-z\\
  \end{array}
 \right),\nonumber\\
&&C=\left(
  \begin{array}{ccc}
 (1+x)\lambda^{-1}-\lambda^{-1}\psi\psi^T\lambda^{-1}&\lambda^{-1}\tilde\omega-J(z-\mu J \lambda^{-1})\psi\\
 \tilde\omega^T\lambda^{-1}+\psi^T(z+\mu \lambda^{-1}J)J&[\tilde\omega^T\lambda^{-1}\tilde\omega-2\mu\psi^T\lambda^{-1}\tilde\omega-\tau(1+x-2y-xy+z^2)]\\
  \end{array}
 \right),\nonumber\\
&&U=
\left(
  \begin{array}{cc}
  (1+x-\mu J\lambda^{-1})\psi-\mu \tau^{-1}\tilde\omega\\
  \mu\tau^{-1}\\
  \end{array}
 \right),\nonumber\\
&&V=\left(
  \begin{array}{ccc}
  (\lambda^{-1}+\mu\tau^{-1}J)\psi\\
  \psi^T\lambda^{-1}\tilde\omega-\mu(1+x-z)\nonumber\\
  \end{array}
 \right),\nonumber\\
&&S=1+2(x-y),\nonumber
\end{eqnarray}
where
\begin{eqnarray}
&&\tilde \omega=\omega-\mu\psi,
\end{eqnarray}
\begin{eqnarray}
&&x=\psi^T\lambda^{-1}\psi,\quad y=\tau^{-1}\mu^2,\quad z=y-\tau^{-1}\psi^TJ\tilde\omega,
\end{eqnarray}
and the matrix $J$ is
\begin{eqnarray}
J= \left(
  \begin{array}{ccccccc}
   0&1\\
   -1&0\\
  \end{array}
 \right).
\end{eqnarray}
$M$ is symmetric ($M^T=M$). $M$ transforms as a symmetric, second-rank tensor field under global $G_{2(2)}$ transformations. The action of $G_{2(2)}$ on the coset matrix can be written as $M'(\Phi^A)={\cal C}^TM(\Phi^A){\cal C}$, where ${\cal C}$ is an exponential of $g_{2(2)}$ Lie algebra generator and $\Phi^A=(\lambda_{ab},\omega_a,\psi_a,\mu)$ expresses the coordinates in the target space.
First, one needs read off the target space variables $\Phi^A$ for a seed solution  via Eqs. (\ref{eq:dual1}) and (\ref{eq:dual2}). The new target space variables $\Phi^{\prime A}=(\lambda'_{ab},\omega'_a,\psi'_a,\mu')$ can be written in terms of the target space variables $\Phi^A$ of a seed from $M(\Phi^{\prime A})=M'(\Phi^A)$. At the final step, by solving the set of  Eqs. (\ref{eq:dual1}) and (\ref{eq:dual2}) for the new variables $\Phi^{\prime A}$, one can obtain the explicit forms of the metric and the gauge potential one-form of a new solution.

\medskip 
In Ref.~\cite{BCCGSW}, they have succeeded in constructing the Harrison transforamtion to preserve asymptotic flatness, which can generate charged solutions to the five-dimensional Einstein-Maxwell-Chern-Simons equations from vacuum solutions to five-dimensional Einstein-equations.
Following Ref.~\cite{BCCGSW},
the transformed coset matrix can be expressed in terms of the seed matrix as
\begin{eqnarray}
M'={\cal C}^TM{\cal C},
\end{eqnarray}
where the one-parameter subgroup is given by
\begin{eqnarray}
{\cal C}=
 \left(
  \begin{array}{ccccccc}
   c^2 & 0 & 0 &s^2 & 0 & 0 &\sqrt{2}sc \\
   0   & c & 0 & 0  & 0 & s &0 \\
   0   & 0 & c & 0  &-s & 0 &0 \\
   s^2 & 0 & 0 & c^2&0 &  0 &\sqrt{2}sc \\
   0   & 0 & -s & 0 &c & 0 &0 \\
   0   & s & 0 & 0  &0 & c &0 \\
\sqrt{2}sc & 0 & 0 & \sqrt{2}sc  &0 & 0 &c^2+s^2 \\
  \end{array}
 \right),
\end{eqnarray}
where $c:=\cosh\gamma$ and $s:=\sinh\gamma$, and $\gamma$ is a parameter. Note that when $\gamma=0$, ${\cal C}$ is a $7\times 7$ unit matrix.
From this one can read off transformed potentials as follows
\begin{eqnarray}
&&\lambda_{00}'=D^{-2}\lambda_{00},\\
&&\lambda_{01}'=D^{-2}(c^3\lambda_{01}+s^3\lambda_{00}\omega_0),\\
&&\tau'=D^{-1}\tau,\\
&&\omega_0'=D^{-2}[c^3(c^2+s^2+2s^2\lambda_{00})\omega_0-s^3(2c^2+(c^2+s^2)\lambda_{00})\lambda_{01}],\\
&&\omega_1'=\omega_1+D^{-2}s^3[-c^3\lambda_{01}^2+s(2c^2-\lambda_{00})\lambda_{01}\omega_0-c^3\omega_0^2],\\
&&\psi_0'=D^{-1}sc(1+\lambda_{00}),\\
&&\psi_1'=D^{-1}sc(c\lambda_{01}-s\omega_0),\\
&&\mu'=D^{-1}sc(c\omega_0-s\lambda_{01}),
\end{eqnarray}
where
\begin{eqnarray}
&&D=1+s^2(1+\lambda_{00}).
\end{eqnarray}

\section{Rasheed solutions}\label{sec:Rasheed}
In the next section, using the $G_{2(2)}$ transforamtion mentioned in the previous section, we will generate a new charged rotating black hole solution from the Resheed solution~\cite{Rasheed}, which the most general Kaluza-Klein black hole solution to the five-dimensional vacuum Einstein equations. Hence, in this section, we shortly explain the solution.

\subsection{Metric}
As shown by Maison~\cite{Maison}, assuming symmetry of a space-time, the five-dimensional pure Einstein equations reduces to equations for five scalar fields on a three dimensional space. The system of the scalar fields is descrived by a nonlinear $\sigma$ model which is invariant under grobal $SL(3,{\bm R})$ transformation. 
In~\cite{Rasheed}, Rasheed derived the black hole solution by applying the $SL(3,{\bm R})$ transformation to a trivial $\rm S^1$ bundle over the the Kerr space-time with the mass parameter $M_k$ and the rotation parameter $a$. Actually to assure the asymptotic flatness in the context of four dimensions, the $SL(3,{\bm R})$ transformation is restricted to the restricted $SO(1,2)$ transformation which are labeled by two boost parameters $(\alpha,\beta)$. The generated new Kalzua-Klein black hole solution is specified by boost parameters $(\alpha,\beta)$ in addition to old parameters $(M_k,a)$. These parameters, $\alpha$ and $\beta$, are related to an electric charge and a magnetric charge under dimensional reduction, restrictively.  
 
\medskip
The metric is given by
\begin{eqnarray}
ds^2&=&\frac{B}{A}(dx^5+2A_\mu dx^\mu)^2-\frac{f^2}{B}(dt+\omega^0{}_\phi d\phi)^2\nonumber\\
    & &+\frac{A}{\Delta}dr^2+Ad\theta^2+\frac{A\Delta}{f^2}\sin^2\theta d\phi^2,\label{eq:Rasheed}
\end{eqnarray}
where
\begin{eqnarray}
&&A=\left(r-\frac{\Sigma}{\sqrt{3}}\right)^2-\frac{2P^2\Sigma}{\Sigma-\sqrt{3}M}+a^2\cos^2\theta+\frac{2JPQ\cos\theta}{(M+\Sigma/\sqrt{3})^2-Q^2},\\
&&B=\left(r+\frac{\Sigma}{\sqrt{3}}\right)^2-\frac{2Q^2\Sigma}{\Sigma+\sqrt{3}M}+a^2\cos^2\theta-\frac{2JPQ\cos\theta}{(M-\Sigma/\sqrt{3})^2-P^2},\\
&&C=2Q(r-\Sigma/\sqrt{3})-\frac{2PJ\cos\theta(M+\Sigma/\sqrt{3})}{(M-\Sigma/\sqrt{3})^2-P^2},\\
&&\omega^0{}_\phi=\frac{2J\sin^2\theta}{f^2}\left[r-M+\frac{(M^2+\Sigma^2-P^2-Q^2)(M+\Sigma/\sqrt{3})}{(M+\Sigma/\sqrt{3})^2-Q^2}\right],\\
&&\omega^5{}_\phi=\frac{2P\Delta \cos\theta}{f^2}-\frac{2QJ\sin^2\theta[r(M-\Sigma/\sqrt{3})+M\Sigma/\sqrt{3}+\Sigma^2-P^2-Q^2]}{f^2[(M+\Sigma/\sqrt{3})^2-Q^2]},\\
&&\Delta=r^2-2Mr+P^2+Q^2-\Sigma^2+a^2,\\
&&f^2=r^2-2Mr +P^2+Q^2-\Sigma^2+a^2\cos^2\theta,\\
&&2A_\mu dx^\mu=\frac{C}{B}dt+\left(\omega^5{}_\phi+\frac{C}{B}\omega^0{}_\phi\right)d\phi,
\end{eqnarray}
where $A_\mu$ is the electromagnatic vector potential derived by dimensional reduction.
The parameters are parameterized by the boost parameters $(\alpha,\beta)$ as
\begin{eqnarray}
&&M=\frac{(1+\cosh^2\alpha\cosh^2\beta)\cosh\alpha}{2\sqrt{1+\sinh^2\alpha\cosh^2\beta}}M_k,\\
&&\Sigma=\frac{\sqrt{3}\cosh\alpha(1-\cosh^2\beta+\sinh^2\alpha\cosh^2\beta)}{2\sqrt{1+\sinh^2\alpha\cosh^2\beta}}M_k,\\
&&Q=\sinh\alpha\sqrt{1+\sinh^2\alpha\cosh^2\beta}\ M_k,\\
&&P=\frac{\sinh\beta \cosh\beta}{\sqrt{1+\sinh^2\alpha\cosh^2\beta}}M_k,\\
&&J= \cosh\beta \sqrt{1+\sinh^2\alpha\cosh^2\beta\ }aM_k,
\end{eqnarray}
where $M,P,Q,J,\Sigma$ mean the mass, the magnatic charge, the electric charge, the angular momentum and dilaton charge, respectively. It should be noted that these parameters are not independent since they are related the equation
\begin{eqnarray}
\frac{Q^2}{\Sigma+\sqrt{3}M}+\frac{P^2}{\Sigma-\sqrt{3}M}=\frac{2\Sigma}{3},
\end{eqnarray}
and $M_k$ is written in terms of these parameters as
\begin{eqnarray}
M_k^2=M^2+\Sigma^2-P^2-Q^2.
\end{eqnarray}
$J$ are also related to $a$ by
\begin{eqnarray}
J^2=a^2\frac{[(M+\Sigma/\sqrt{3})^2-Q^2][(M-\Sigma/\sqrt{3})^2-P^2]}{M^2+\Sigma^2-P^2-Q^2}.
\end{eqnarray}
 The locations of the horizon is given by the zeros of $\Delta$ when the parameters satisfy the inequality
\begin{eqnarray}
M^2\ge P^2+Q^2+a^2-\Sigma^2.
\end{eqnarray}

\subsection{Potentials}
First, one has to indentify the target space variables $\Phi^A=(\lambda_{ab},\omega_a,\psi_a,\mu)$ which is corresponds to the seed solution.
Choose the coordinates as $z^0=t,z^1=x^5,x^2=\phi,x^3=r$ and $x^4=\theta$. In this paper, we restrict the boost parameters to $\alpha=0$ for simplicity. Under these identification, we can read off the target space variables $\Phi^A$ from the metric (\ref{eq:Rasheed}) as follows
\begin{eqnarray}
&&\lambda_{00}=g_{tt}=-\frac{f^2}{B}+4\frac{B}{A}A_t^2,\\
&&\lambda_{01}=g_{t5}=2\frac{B}{A}A_t,\\
&&\lambda_{11}=g_{55}=\frac{B}{A},\\
&&\omega_0=\frac{2aM_k\cos\theta \cosh\beta}{A},\\
&&\omega_1=\frac{M_k(2r-M_k\sinh^2\beta)\sinh\beta\cosh\beta}{A},\\
&&\tau=-{\rm det\ (\lambda}_{ab})=\frac{f^2}{A},\\
&&\psi_0=0,\\
&&\psi_1=0,\\
&&\mu=0,
\end{eqnarray}
where in order to obtain $\omega_a\ (a=0,1)$, we substituted the equations
\begin{eqnarray}
&&a^0{}_\phi=\frac{\lambda_{01}g_{\phi5}-\lambda_{11}g_{t\phi}}{\tau}=\omega^0{}_\phi,\\
&&a^1{}_\phi=-\frac{\lambda_{00}g_{\phi5}-\lambda_{01}g_{t\phi}}{\tau}=\omega^5{}_\phi
\end{eqnarray}
into Eqs.(\ref{eq:dual1}) and (\ref{eq:dual2}) and integrated them. The three-dimensional metric $h_{ij}$ and the square root of the determinant $h$ are 
\begin{eqnarray}
&&h_{rr}=\tau g_{rr}=\frac{f^2}{\Delta},\\
&&h_{\theta\theta}=\tau g_{\theta\theta}=f^2,\\
&&h_{\phi\phi}=\tau (g_{\phi\phi}-2\lambda_{01}a^0{}_\phi a^1{}_\phi-\lambda_{11}(a^1{}_\phi)^2-\lambda_{00}(a^0{}_\phi)^2)=\Delta \sin^2\theta,\\
&&\sqrt{h}=f^2\sin\theta.
\end{eqnarray}

\section{New charged solutions}\label{sec:new}

\subsection{New solution}
To obtain the metric and the gauge potentail one-form of the Maxwell field of the new solution, one must solve Eqs.(\ref{eq:dual1}) and (\ref{eq:dual2}) for the transformed target space variables $\Phi^{\prime A}=(\lambda'_{ab},\omega'_a,\psi'_a,\mu')$. Integrating them, we can obrain the explicit forms $a^{\prime a}{}_\phi$ and $A_\phi$ for the new solution, respectively,  as follows
\begin{eqnarray}
&&a^{\prime 0}{}_\phi=c^3a^{ 0}{}_\phi+s^3 y,\nonumber\\
&&a^{\prime 1}{}_\phi=a^{ 1}{}_\phi,\nonumber\\
&&A_\phi=\frac{-2\sqrt{3}scaM_k[\sin^2\theta \cosh(\beta+\gamma)(r+3M_k\sinh^2\beta/2)+\cos^2\theta\sinh(\beta+\gamma)M_k\sinh2\beta]d\phi}{(r+3M_k\sinh^2\beta/2)(r+2s^2M_k-M_k\sinh^2\beta/2)+a^2\cos^2\theta},\nonumber
\end{eqnarray} 
where the function $y$ is given by
\begin{eqnarray}
&&y=\frac{a M_k(2r-M_k\sinh^2\beta-4M_k)\sin^2\theta \sinh\beta}{f^2}.
\end{eqnarray}
Then, in terms of old varibles,
the metric generated by $G_{2(2)}$ transformation can be expressed in the form 
\begin{eqnarray}
ds^2&=&D^{-2}\lambda_{00}[dt+\bm \Omega']^2\nonumber\\
   &&+D\biggl[\tau^{-1}h_{\phi\phi}d\phi^2-\frac{\tau}{\lambda_{00}}\left(dx^5+a^1{}_\phi d\phi\right)^2+\tau^{-1}(h_{rr}dr^2+h_{\theta\theta} d\theta^2)\biggr],\label{eq:newD}
\end{eqnarray}
where the one-form $\bm\Omega'$ can be written as follows
\begin{eqnarray}
&&\bm\Omega'=\Omega'_\phi d\phi+\Omega'_5dx^5
\end{eqnarray}
with
\begin{eqnarray}
&&\Omega'_\phi=c^3\Omega_\phi+s^3(y+\omega_0a^1{}_\phi),\\
&&\Omega'_5=c^3\Omega_5+s^3\omega_0,\\
&&\Omega_\phi=a^0{}_\phi+\frac{\lambda_{01}}{\lambda_{00}}a^1{}_\phi,\\
&&\Omega_5=\frac{\lambda_{01}}{\lambda_{00}}.
\end{eqnarray}

\medskip
Here we introduce a new radial coordinate $\rho:=r+2s^2M_k-\frac{1}{2}\sinh^2\beta M_k$. Then in terms of the coordinates $(t,x^5,\phi,\rho,\theta)$ and parameters including the seed solution, the metric can be explicitly rewritten as
\begin{eqnarray}
ds^2&=&-\frac{XY}{Z^2}[dt+{\bm \Omega'}]^2+\frac{Z}{Y}\Biggl[\frac{W}{X}\left(dx^5+\frac{2P\Delta \cos\theta}{W}d\phi\right)^2\nonumber\\
&&+Y\left(\frac{\Delta\sin^2\theta}{W} d\phi^2+\frac{d\rho^2}{\Delta}+d\theta^2\right)\Biggr],
\end{eqnarray}
where the functions $X,Y,Z,W$ and the one-form $\bm\Omega'$ are given by
\begin{eqnarray}
&&X=(\rho-2c^2M_k)(\rho-2s^2M_k+2M_k\sinh^2\beta)+a^2\cos^2\theta, \\
&&Y=(\rho-2s^2M_k)(\rho-2s^2M_k+2M_k\sinh^2\beta)+a^2\cos^2\theta, \\
&&Z=\rho(\rho-2s^2M_k+2M_k\sinh^2\beta)+a^2\cos^2\theta,\\
&&W=(\rho-2c^2M_k)(\rho-2s^2M_k)+a^2\cos^2\theta,\\
&&\Delta=(\rho-2c^2M_k)(\rho-2s^2M_k)+a^2
\end{eqnarray}
and
\begin{eqnarray}
{\bm \Omega'}=\left(c^3\frac{X_\phi}{X}+s^3\frac{Y_\phi}{Y}\right)d\phi+\left(c^3\frac{X_5}{X}+s^3\frac{Y_5}{Y}\right)dx^5
\end{eqnarray}
with
\begin{eqnarray}
&&X_\phi=2aM_k\cosh\beta[\sin^2\theta(\rho-2M_ks^2)+2M_k\sinh^2\beta],\\
&&X_5=2aM_k\cos\theta\sinh\beta,\\
&&Y_\phi=2aM_k\sinh\beta[\sin^2\theta(\rho-2M_kc^2)+2M_k\cosh^2\beta],\\
&&Y_5=2aM_k\cos\theta\cosh\beta.
\end{eqnarray}
The gauge potential one-form of the Maxwell field is given by
\begin{eqnarray}
{\bm A}&=&\frac{2\sqrt{3}scM_k}{Z}\biggl[[\rho+2M_k(\sinh^2\beta-s^2)]dt+a\cos\theta\sinh(\beta+\gamma)dx^5\nonumber\\
&-&a\left(\sin^2\theta\cosh(\beta+\gamma)(\rho+2M_k(\sinh^2\beta-s^2))+\cos^2\theta \sinh(\beta+\gamma)\sinh2\beta M_k \right)d\phi\biggr].
\end{eqnarray}

This new solution has four independet parameters $(M_k,a,\beta,\gamma)$. The parameter region such that there are two black hole horizons is  
\begin{eqnarray}
M_k>0,\quad M_k^2>a^2. 
\end{eqnarray} 
As shown later, in the sense of five dimensions, they are physically related to the mass, the angular momentum in the direction of four dimensions, the charge and the size of an extra dimension at infinity. Instead of $(M_k,\gamma)$, it is better to use the physical parameters $(m,q)$ which are proportional to the mass and the charge as follows
\begin{eqnarray}
m=(s^2+c^2)M_k,\quad q=2scM_k,
\end{eqnarray}
but in this paper we do not use these to avoid the complicated form of the metric.

\section{Basic Properties}\label{sec:property}

\subsection{Asymptotic structure and asymptotic charges}
At the infinity, $\rho \to \infty$, the metric behaves as
\begin{eqnarray}
ds^2 \simeq -dt^2+d\rho^2+\rho^2(d\theta^2+\sin^2\theta d\phi^2)+(dx^5+2P \cos\theta d\phi)^2.
\end{eqnarray}
The space-time is asymptotically locally flat, in particular, in the case of $P\not =0$, the space-time asymptotically approaches a twisted $\rm S^1$ bundle over a four-dimensional Minkowski space-time. We set the perioodicity of $x^5$ to $8\pi P$. Then the spatial infinity is a squashed $\rm S^3$. In the case of $P=0$, the asymptotic structure is a trivial $\rm S^1$ bundle over a four-dimensional Minkowski space-time.
 From Eq.(\ref{eq:newD}), asymptotic structure is preserved since $D \to 1$ and $\bm\Omega'\to 0$ at infinity.

\medskip
We give the conserved charges of the new solution. 
The charge is
\begin{eqnarray}
{\bm Q_e}=\frac{1}{4\pi G_5}\int\left(* F-\frac{1}{\sqrt{3}}A\wedge F\right)=-\frac{16\sqrt{3}\pi}{G_5}  \sinh\beta\cosh\beta M_k^2sc.
\end{eqnarray}
The Komar mass and the Komar angular momenta at infinity, which are associated with Killing vector fields $t^\mu=(\partial_t)^\mu$, $\phi^\mu=(\partial_\phi)^\mu$ and $\psi^\mu=(\partial_{\psi})^\mu$, is
\begin{eqnarray}
&&M^{\rm Komar}=-\frac{3}{32\pi G_5}\int_{\rm S_\infty}dS_{\mu\nu}\nabla^\mu t^\nu=\frac{6\pi}{G_5} \sinh\beta\cosh\beta(c^2+s^2)M_k^2,\\
&&J^{\rm Komar}_\phi=-\frac{1}{16\pi G_5}\int_{\rm S_\infty}dS_{\mu\nu}\nabla^\mu\phi^\nu=-\frac{8\pi}{3G_5} \sinh\beta\cosh\beta (c^3\cosh\beta+s^3\sinh\beta)M_k^2a,\\
&&J^{\rm Komar}_\psi=-\frac{1}{16\pi G_5}\int_{\rm S_\infty}dS_{\mu\nu}\nabla^\mu\psi^\nu=0,
\end{eqnarray}
respectively. Here the angular coordinate $\psi$ with periodicity of $4\pi$ is defined by $\psi:=x^5/2P$. Therefore, the space-time has only an angular momentum in the direction of four-dimensions.

\subsection{Horizon and regularity}
Two black hole horizons, an outer horizon and an inner horizon,  are located at $\rho=\rho_\pm\ (\rho_-<\rho_+)$, where $\rho_\pm$ is roots of the quadratic equation, $\Delta=0$, with respect to $\rho$ and they are explicitly written in terms of the parameters $M_k,a,\gamma$ as
\begin{eqnarray}
\rho_\pm:=M_k+2s^2M_k\pm \sqrt{M_k^2-a^2}.
\end{eqnarray}
The functions $X,Y,Z,W,\Omega'_\phi$ and $\Omega_5'$ do not vanish at $\rho=\rho_\pm$ and only the metric component $g_{\rho\rho}$ apparently diverges there.
In order to remove the divergence, we introduce the coordinates $(t',\phi')$ defined as
\begin{eqnarray}
&&dt'=dt-\frac{a }{2\Omega_H\sqrt{M_k^2-a^2}}\frac{d\rho}{\rho-\rho_+},\\
&&d\phi'=d\phi-\frac{a }{2\sqrt{M_k^2-a^2}}\frac{d\rho}{\rho-\rho_+},
\end{eqnarray}
where $\Omega_H$ denotes the angular velocity of the horizon and is given by 
\begin{eqnarray}
\Omega_{H}&=&-\frac{1}{\Omega'_\phi(\rho=\rho_+)}\nonumber\\
&=&\frac{a}{2M_k[c^3(M_k+\sqrt{M_k^2-a^2})\cosh\beta+s^3(-M_k+\sqrt{M_k^2-a^2})\sinh\beta]}.
\end{eqnarray}
The black hole is rotating only in the direction of $\partial_\phi$.
Then near $\rho=\rho_+$, the metric behaves as
\begin{eqnarray}
ds^2&\simeq&-\frac{X_+Y_+}{Z_+^2}[dt'+\Omega'_{\phi+} d\phi'+\Omega'_{5+} dx^5]^2+\frac{Z_+}{Y_+}\Biggl[\frac{W_+}{X_+}\left( dx^5+\frac{2Pa \cos\theta }{W_+} d\rho \right)^2\nonumber\\
   & &+2\frac{aY_+\sin^2\theta}{W_+}d\phi'd\rho +Y_+d\theta^2\Biggr]+{\cal O}(\rho-\rho_+),
\end{eqnarray}
where $(X_+,Y_+,Z_+,W_+,\Omega_{5+},\Omega_{\phi+})$ denote the values on the $\rho=\rho_+$ of the functions $(X,Y,Z,W,\Omega_{5},\Omega_{\phi})$.
The metric is well-defined on $\rho=\rho_+$.
Furthermore, to show that $\rho=\rho_+$ corresponds to the horizon, introduce new coordinates $(u,\phi'')$ defined as
\begin{eqnarray}
&&du=dt',\\
&&d\phi''=d\phi'-\Omega_H dt'.
\end{eqnarray}
Then, near $\rho=\rho_+$, the metric behaves as
\begin{eqnarray}
ds^2&\simeq& -\frac{X_+Y_+}{Z_+^2}[\Omega'_{\phi+} d\phi''+\Omega'_{5+} dx^5]^2+\frac{Z_+}{Y_+}\Biggl[\frac{W_+}{X_+}\left( dx^5+\frac{2Pa \cos\theta }{W_+} d\rho \right)^2\nonumber\\
   & &+2\frac{a Y_+ \sin^2\theta}{W_+}d\phi''d\rho +Y_+d\theta^2\Biggr]+\frac{2aZ_+\Omega_H \sin^2\theta}{W_+}du d\rho+{\cal O}(\rho-\rho_+).
\end{eqnarray}
The Killing vector field $V:=\partial_u$ is null on $\rho=\rho_+$. Since $V_\mu dx^\mu=g_{\rho u}d\rho$  there, the vector field is  orthogonal and tangent to the null surface $\rho=\rho_+$. Hence the null hypersurface $\rho=\rho_+$ is a Killing horizon. In addition, in the coordinate system $(u,x^5,\phi'',\rho,\theta)$, the metric is smooth everywhere in the region $\rho\ge \rho_+$.  So there is no singularity on and outside the horizon. Similarily, the null surface $\rho=\rho_-$ is a Killing horizon.
 
\medskip
The induced metric on the outer horizon is
\begin{eqnarray}
ds^2_{\cal H}&=&\frac{X_+^2Y_+^2\Omega_{5 +}^{\prime 2}-Z_+^3W_+}{-X_+Y_+Z_+^2}\left(dx^5+{\cal A} \right)^2+Z_+\Biggl[d\theta^2-\frac{X_+Y_+W_+\Omega_{\phi+}^{\prime 2}}{Z_+^3W_+-(X_+Y_+\Omega_{5+}^{\prime })^2}d\phi''{}^2\Biggr],\label{eq:horizon}
\end{eqnarray}
where the one-form ${\cal A}$ is given by
\begin{eqnarray}
{\cal A}=-\frac{X_+^2Y_+^2\Omega'_{\phi+}\Omega'_{5+}}{Z_+^3W_+-(X_+Y_+\Omega_{5+}^{\prime })^2}d\phi''.
\end{eqnarray}
The metric can be regarded as a twisted $\rm S^1$ fiber bundle over an $\rm S^2$ base space. Near $\theta=0$ and $\theta=\pi$, the metric on the $\rm S^2$ behaves as
\begin{eqnarray}
&&ds^2\simeq \rho_+(\rho_+-2s^2M_k+2\sinh^2\beta M_k)[d\theta^2+\theta^2 d\phi''{}^2+{\cal O}(\theta^3)],\\
&&ds^2\simeq \rho_+(\rho_+-2s^2M_k+2\sinh^2\beta M_k)[d\theta^2+(\theta-\pi)^2d\phi''{}^2+{\cal O}((\theta-\pi)^3)],
\end{eqnarray}
respectively. Hence there is no conical singularity on the base space. Moreover, in the case of $\beta\not=0$, the first Chern number is computed as
\begin{eqnarray}
c_1=\frac{1}{8\pi P}\int_{S^2}{\cal F}=\frac{{\cal A}_\phi(\theta=\pi)-{\cal A}_\phi(\theta=0)}{4 P}=-1,
\end{eqnarray}
where ${\cal F}=d{\cal A}$.  Therefore we have a non-trivial bundle, i.e., Eq.(\ref{eq:horizon}) is a metric on $\rm S^3$.

\subsection{Ergoregion}
An ergoregion is the region where $g_{tt}=-XY/Z^2$ becomes positive. We see that $Y>0$ outside the outer horizon. Therefore, the ergosurface is located at the zero of the function $X$, i.e.,
\begin{eqnarray}
\rho=M_k(1+2s^2-\sinh^2\beta)+\sqrt{M_k^2\cosh^4\beta-a^2\cos^2\theta}.
\end{eqnarray}
As far as $a\not= 0$, $\rho>\rho_+$ for arbitrary values of $\theta$. The ergosurface is located outside the outer horizon.

\subsection{Various limits}

\subsubsection{$a\to 0$}
Taking the limit of $a\to 0$ and defining the parameters as 
\begin{eqnarray}
&&r_\infty:=4P,\quad \rho_+:=2c^2M_k,\quad \rho_-:=2s^2M_k,\quad \rho_0:=2M_k(\sinh^2\beta-s^2),
\end{eqnarray}
we can obtain the following metric
\begin{eqnarray}
ds^2=-V(\rho)dt^2+\frac{K^2(\rho)}{V(\rho)}d\rho^2+\rho^2K^2(\rho)(d\theta^2+\sin^2\theta d\phi^2)+W^2(\rho)\chi^2,
\end{eqnarray}
where the functions $V,K,W$ are 
\begin{eqnarray}
&&V(\rho)=\frac{(\rho-\rho_-)(\rho-\rho_+)}{\rho^2},\\
&&K^2(\rho)=\frac{\rho+\rho_0}{\rho},\\
&&W^2(\rho)=\frac{r_\infty^2}{4}K^{-2}(\rho).
\end{eqnarray}
This solution coincides with the static charged 
Ishihara-Matsuno solution to the five-dimensional 
Einstein-Maxwell equations. In particular, in the case of 
$r_\infty\to \infty$, the solution becomes 
the five-dimensional Reissner-Nordstr\"om solution, which is 
asymptotically flat in the standard five-dimensional sense.  
In the case of $\rho_+=\rho_-$, it has a degenerate horizon and is supersymmetric 
because it is included in a class of solutions on Taub-NUT base space 
in \cite{Gauntlett}, in which all purely bosonic supersymmetric solutions 
of minimal supergravity in five dimensions are classified.

\subsubsection{$\gamma\to 0$}
In the limit of $\gamma\to 0$, it is clear that $D\to 1$ and $\bm\Omega'\to \bm\Omega$. Therefore this solution coincides with uncharged Resheed solution with $Q=0\ (\alpha=0)$.

\subsubsection{$\beta\to 0$}
In the limit of $\beta\to 0$, $X=W$ and $X_5=0$ and $Y_\phi=0$. Then the metric takes the form of
\begin{eqnarray}
ds^2&=&-\frac{[(\rho-2c^2M_k)(\rho-2s^2M_k)+a^2\cos^2\theta][(\rho-2s^2M_k)^2
+a^2\cos^2\theta]}{[\rho(\rho-2s^2M_k)+a^2\cos^2\theta]^2}(dt+{\bm \Omega'})^2\nonumber\\
    &&+[\rho(\rho-2s^2M_k)+a^2\cos^2\theta]\left[d\theta^2+\frac{\Delta \sin^2\theta d\phi^2}{(\rho-2c^2M_k)(\rho-2s^2M_k)+a^2\cos^2\theta}+\frac{d\rho^2}{\Delta}\right]\nonumber\\
    & &+\frac{\rho(\rho-2s^2M_k)+a^2\cos^2\theta}{(\rho-2s^2M_k)^2+a^2\cos^2\theta}(dx^5)^2,
\end{eqnarray}
where the one-form $\bm \Omega'$ is given by
\begin{eqnarray}
\bm \Omega'=2aM_k\left(\frac{c^3(\rho-2s^2M_k)\sin^2\theta}{(\rho-2c^2M_k)(\rho-2s^2M_k)+a^2\cos^2\theta}d\phi+\frac{s^3\cos\theta}{(\rho-2s^2M_k)^2
+a^2\cos^2\theta}dx^5\right).
\end{eqnarray}
The gauge potential one-form is given by
\begin{eqnarray}
{\bm A}&=&\frac{2\sqrt{3}scM_k\left[(\rho-2s^2M_k)dt+as\cos\theta dx^5
-ac\sin^2\theta (\rho-2s^2M_k)d\phi\right]}{\rho(\rho-2s^2M_k)+a^2\cos^2\theta}.
\end{eqnarray}

\medskip
For the infinity, $\rho \to \infty$, the metric asymptotically behaves as
\begin{eqnarray}
ds^2=-dt^2+d\rho^2+\rho^2(d\theta^2+\sin^2\theta d\phi^2)+(dx^5)^2,
\end{eqnarray}
which is the tivial $\rm S^1$ bundle over a four-dimensional Minkowsky space-time. 
 
\medskip
 Note that ${\cal A}_\phi(\theta=\pi)={\cal A}_\phi(\theta=0)=0$ on the horizon $\rho=\rho_+$. So the first Chern number is
\begin{eqnarray}
c_1=\frac{1}{\Delta x^5}\int_{S^2}{\cal F}=2\pi\frac{{\cal A}_\phi(\theta=\pi)-{\cal A}_\phi(\theta=0)}{\Delta x^5}=0.
\end{eqnarray}
 Therefore the horizon is also a trivial bundle, i.e., it is topologically $\rm S^2 \times S^1$. This solution corresponds to electrically charged rotating black string solution.

\subsubsection{$\beta\to \infty$}
We take the limit of $\beta \to \infty$ with the other variables and the following quanties kept fixed
\begin{eqnarray}
&&\tilde m=8(s^2+c^2)\sinh^2\beta M_k^2,\\ 
&&\tilde q= 16cs\sinh^2\beta M_k^2, \\
&&\tilde a= -2\sinh\beta(c-s)^{-1}a.
\end{eqnarray}
The functions $X,Y,Z$ behaves as the following functions, respectively,
\begin{eqnarray}
&&X\to 2M_k\sinh^2\beta (\rho-2c^2 M_k),\\
&&Y\to 2M_k\sinh^2\beta (\rho-2s^2 M_k),\\
&&Z\to 2M_k\sinh^2\beta\ \rho.
\end{eqnarray} 
Define a new radial coordinate $r$, three new angular coordinates $(\Theta,\phi_1,\phi_2)$ as $r^2=\tilde \rho^2-a^2$ ($\tilde \rho^2=(8M_k\sinh^2\beta) \rho$), $\Theta=\theta/2$, $\phi_1=(\psi-\phi)/2$ and  $\phi_2=(\psi+\phi)/2$, respectively. Noting that the one-form ${\bm \Omega'}$ behaves as
\begin{eqnarray}
{\bm \Omega'}\to -\frac{\tilde a}{2}\frac{(2\tilde m-\tilde q)\tilde\rho^2-\tilde q^2}{\tilde\rho^4-2\tilde m\tilde \rho^2+\tilde q^2}(d\phi+\cos\theta d\psi),
\end{eqnarray}
we obtain the solution
\begin{eqnarray}
ds^2&=&-dt^2-\frac{2\tilde q}{\tilde \rho^2}\tilde \nu(dt-\tilde \omega)+\frac{\tilde f}{\tilde\rho^4}(dt-\tilde\omega)^2+\frac{\tilde \rho^2 r^2}{\tilde \Delta}dr^2\\
&&+\tilde\rho^2[d\Theta^2+\sin^2\Theta d\phi_1^2+\cos^2\Theta d\phi_2^2],
\end{eqnarray}
\begin{eqnarray}
{\bm A}=\frac{\sqrt{3}\tilde q}{\tilde \rho^2}(dt-\tilde\omega),
\end{eqnarray}
where
\begin{eqnarray}
&&\tilde \omega=-\tilde\nu=\tilde a(\sin^2\Theta d\phi_1-\cos^2\Theta d\phi_2),\\
&&\tilde f=2\tilde m\tilde \rho^2 -\tilde q^2,\\
&&\tilde \Delta =(r^2+\tilde a^2)^2+\tilde q^2-2\tilde a^2\tilde q-2\tilde m r^2.
\end{eqnarray}
This solution coincides with the Cvetic-L\"u-Pop solution in Ref.\cite{CLP} for $g=0$, $b=-a$, which is the asymptotically flat charged black hole solution with equal angular momenta with an opposite sign $J_{\phi_1}=-J_{\phi_2}$.

\subsubsection{$M_k\to a$}
According to \cite{Gauntlett}, all supersymmetric 
solutions of the five-dimensional minimal supergravity have a non-spacelike 
Killing vector field, and 
when the Killing vector field $\partial_t$ is timelike,
the metric and the gauge potential are given, respectively, by 
\begin{eqnarray}
ds^2=-H^{-2}(dt+\bm\omega)^2+Hds^2_{{\cal B}} \,, 
\ {\bm A}=\frac{\sqrt{3}}{2}[H^{-1}(dt+\bm\omega)-\bm\beta] \,, 
\label{eq:SUSY}
\end{eqnarray}
where $ds^2_{\cal B}$ is a metric of a hyper-K\"ahler space ${\cal B}$. 
The scalar function $H$, one-forms $\bm\omega$ and $\bm\beta$ on ${\cal B}$ 
are given by 
\begin{eqnarray}
\Delta H=\frac{4}{9}(G^+)^2 \,,\quad 
dG^+=0 \,,\quad 
d{\bm \beta}=\frac{2}{3}G^+ \,. 
\label{eq:SUSY2} 
\end{eqnarray}
Here, $\triangle$ is the Laplacian on ${\cal B}$ and the two-form $G^+$ is 
the self-dual part of the one-form $H^{-1}{\bm \omega}$, given by 
\begin{eqnarray}
  G^+:=\frac{1}{2}H^{-1}(d\omega+*d\omega) \,,  
\label{eq:SUSY3}  
\end{eqnarray}
where $(G^+)^2:=\frac{1}{2}G_{mn}G^{mn}$ and $*$ is the Hodge dual operator on 
${\cal B}$. Since $\partial_t$ is a Killing vector field associated 
with time translation, all components are independent of 
the time coordinate $t$.

When the parameters satisfies $M_k=a$ in our solution, two horizons degenerate. However, there is no timelike Killing vector field which grobally exsits outside the outer horizon except the case of $a=0$. Therefore this is non-supersymmetric. In the case of $a=0$,  $\rho_+=\rho_-$ corresponds to the extreme and supersymmetric case. Introducing the coordinate $R:=\rho-\rho_+$ and the parameter $N:=\rho_0+\rho_+$, we obtain the metric
\begin{eqnarray}
&&ds^2=H^{-2}dt^2+Hds_{TN}^2,\\
&&H=1+\frac{\rho_+}{R},\\
&&ds^2_{TN}=\left(1+\frac{N}{R}\right)^{-1}(dx^5+\cos\theta d\phi)^2+\left(1+\frac{N}{R}\right)(dR^2+R^2d\theta^2+R^2\sin^2\theta d\phi^2).
\end{eqnarray}
$ds^2_{TN}$ is the metric of a self-dual Euclidean Taun-NUT space. Hence this is supersymmetric black hole solutions on Taub-NUT space~\cite{Gauntlett}.

\section{Summary and discussions}\label{sec:summary}
In this paper, applying $G_{2(2)}$ generating techniques for minimal $D=5$ supergravity to the Rasheed Kalzua-Klein black hole solution~\cite{Rasheed} with the restricted case $\alpha=0$, we have obtained a new charged rotating Kaluza-Klein black hole solution to the five-dimensional Einstein-Maxwell-Chern-Simons equations. The solution has four independent parameters $(M_k,a,\beta,\gamma)$. The solution is a generalization of the Ishihara-Matsuno static charged solution~\cite{IM} to the rotating case. The space-time has two black hole horizons, an outer horizon and an inner horizon. The event horizon on the cross sections with timeslice is topologically $\rm S^3$. Thus, near the horizon, the space-time behaves as five dimensions but at infinity the space-time effectively behaves as four dimensional space-time, which has a twisted $\rm S^1$ bundle over a four dimensional Miskowsky space-time.
Our new solution has many limits to several known solutions, e.g., (i) $\beta= \infty$ case: the asymptotically flat charged rotating black hole with two equal angular momenta~\cite{CLP}, (ii) $a=0$ case: the Ishihara-Matsuno static charged Kalzua-Klein black hole solution~\cite{IM}, and (iii) $\beta=0$ case: a new electrically charged rotating black string solution which asymptotes to a trivial bundle over four dimensional Miskowsky space-time.

\medskip
It is interesting to compare our solution with the rotaitng solution in~\cite{NIMT}. These two solutions are the generalization of Ishihara-Matsuno static charged black hole solution~\cite{IM} to rotating black hole solutions.
As mentioned previousely, the extreme case (i.e., the $m=\pm q$ case) of the Ishihara-Matsuno solution corresponds to a supersymmetric solution, which is a black hole solution on the Euclidean self-dual Taub-NUT space~\cite{Gauntlett}. 
However, in the solution in~\cite{NIMT}, the two cases, $m=-q$ and $m=q$, describe different solutions due to the existence of a Chern-Simons term. 
Only the case of $m=-q$ corresponds to a supersymmetric Kaluza-Klein black 
hole solution which was found by Gaiotto {\it et al.}~\cite{Gaiotto}. In our solution, two extreme cases ($M_k^2=a^2,\gamma>0$) and ($M_k^2=a^2,\gamma<0$) also become different solutions by the existence of a Chern-Simons term. Furthermore, in contrast to the solution in~\cite{NIMT}, neither ($M_k^2=a^2,\gamma>0$) nor ($M_k^2=a^2,\gamma<0$) is a supersymmetric solution.

\medskip
In this paper, we restricted the Rasheed solution to the special case of $\alpha=0$ since it is more difficult to solve Eqs.(\ref{eq:dual1}) and (\ref{eq:dual2}) generally. The solution generated from the general seed solution with $\alpha\not =0$ and $\beta\not= 0$ by the $G_{2(2)}$ transformation will include both our solution and the solution in~\cite{NIMT}, i.e., in such a solution, black hole will be rotaing in the extra dimension $\partial_{x^5}$ as well as in the direction of $\partial_\phi$. The construction of such a solution is our future work.

\medskip
Finally, we would like to comment on the squashing transformation. The squashing transformation is a very powerful tool to generate Kalzua-Klein black hole solutions from asymptotically flat and non-asymptotically flat black hole solutions.  The transforamtion can be applied only to cohomogeneity-one black hole solution such as the five-dimensional Myers-Perry black hole solution~\cite{MP} with two equal angular momenta and the Cvetic-L\"u-Pop black hole solution~\cite{CLP} with two equal angular momenta. The solutions generated by the transformation also belong to a class of cohomogeneity-one Kalzua-Klein black hole solutions~\cite{Wang,NIMT,TIMN,TI}. However, our solution cannot be obtained by the transformarion since our solution belongs a class of cohomogeneity-two, though our solution has the limit to the cohomogeneity-one asymptotically flat Cvetic-L\"u-Pop black hole solution with two equal angular momenta.

\section*{Acknowledgments} 
ST is suppoted by the JSPS under Contract No. 20-10616. YM is supported by the OUEL research fund. YY is supported by the Grant-in Aid for Scientific Research (No. 19540304 and No. 19540098)

\appendix

\end{document}